\documentclass{aa}  
\usepackage{graphicx}
\usepackage{natbib}
\bibpunct{(}{)}{;}{a}{}{,} 
\usepackage{txfonts}
\def\grb11{GRB\ 060111B}
\def\ep{E$_{\mathrm{peak}}$}

\begin{document}
%

  \title{A multiwavelength study of Swift GRB\ 060111B constraining the origin of its prompt optical emission
   \thanks{Based on observations performed with: TAROT at the
   Calern observatory. The Nordic Optical Telescope, operated 
   on the island of La Palma jointly by Denmark, Finland,
   Iceland, Norway, and Sweden, in the Spanish Observatorio
   del Roque de los Muchachos of the Instituto de Astrofisica
   de Canarias. CFHT, which is operated by the National
   Research Council of Canada, the Institut National des Sciences
   de l'Univers of the CNRS of France, and the University of
   Hawaii. The 2.6m Shajn telescope at the Crimean Astrophysical Observatory. Some of the data presented here were obtained at the W.M. Keck Observatory, which is operated as a scientific partnership among the California Institute of Technology, the University of California and the National Aeronautics and Space Administration. The Observatory was made possible by the generous financial support of the W.M. Keck Foundation.}
   }

   \subtitle{}

   \author{G. Stratta\inst{1}
          \and
          A. Pozanenko\inst{2}
          \and
          J-L. Atteia\inst{3}
          \and
          A. Klotz \inst{4}
          \and
          S. Basa\inst{5}
          \and
          B. Gendre\inst{5}
          \and
          F. Verrecchia\inst{1}
          \and
          M. Bo\"er\inst{6}
          \and
	  S. Cutini\inst{1}
	  \and
          M. Henze\inst{7}
          \and
          S. Holland\inst{8}
          \and
	  M. Ibrahimov\inst{9}
	  \and
          F. Ienna\inst{3}
          \and
          I. Khamitov\inst{10}
          \and
          S. Klose\inst{7}
          \and
          V. Rumyantsev\inst{11}
          \and
	  V. Biryukov\inst{12}
	  \and
          D. Sharapov\inst{9}
          \and
          F. Vachier\inst{13}
	  \and
	  S. Arnouts\inst{5,14}
	  \and
	  D.A. Perley\inst{15}
       }

   \offprints{G. Stratta}

   \institute{ASDC, ASI Science Data Center, Via Galileo Galilei, 00044 Frascati, Italy 
         \and
         Space Research Institude IKI, Profsoyznaya 84/32, 117910 Moscow, Russia 
         \and
         LATT, Universit\'e de Toulouse, CNRS,14 Av. E. Belin, F-31400 Toulouse, France 
         \and
         CESR, Universit\'e de Toulouse, CNRS, BP 4346, F-31028 - Toulouse Cedex 04, France 
         \and
         LAM, Observatoire Astronomique de Marseille, P\^ole de l'\'Etoile Site de Ch\^ateau-Gombert 38, rue Fr\'ed\'eric Joliot-Curie 13388 Marseille cedex 13, France 
         \and
         Observatoire de Haute Provence, F-04870 Saint Michel l'Observatoire, France 
         \and
         Th\"ueringer Landessternwarte Tautenburg, Sternwarte 5, D-07778 Tautenburg, Germany 
         \and
         NASA, Goddard Space Flight Center, Greenbelt, MD 20771, USA 
         \and
         Ulugh Beg Astronomical Institute, Tashkent 700052, Uzbekistan  
         \and
         T\"UBITAK National Observatory, Akdeniz \"Universitesi, 07058, Antalya, Turkey 
         \and
         SRI Crimean Astrophysical Observatory, Nauchny, Crimea, 98409, Ukraine 
         \and
	 Crimean Laboratory of the Sternberg Astronomical Institute, Nauchny, Crimea, 98409, Ukraine
	 \and
         IMCCE, Observatoire de Paris, 77 Avenue Denfert Rochereau, F-75014 - Paris, France 
	 \and
	 Canada France Hawaii Telescope  corporation,  65-1238 Mamalahoa Hwy, Kamuela, HI 96743, USA 
	 \and
	 Department of Astronomy, University of California, Berkeley, CA 94720-3411, USA 
         }

   \date{Received ; accepted }

 
 \abstract{
   {The detection of bright optical emission measured with good temporal resolution during the prompt phase of \grb11 makes this GRB a rare event that is especially useful for constraining theories of the prompt emission.    }
   {For this reason an extended multi-wavelength campaign was performed to further constrain the physical interpretation of the observations.  }
   {In this work, we present the results obtained from our multi-wavelength campaign, as well as from the public Swift/BAT, XRT, and UVOT data.}
   {We identified the host galaxy at $R\sim25$ mag from deep $R$-band exposures taken 5 months after the trigger. Its featureless spectrum and brightness, as well as the non-detection of any associated supernova 16 days after the trigger, enabled us to constrain the distance scale of \grb11 within $0.4\le z \le3$ in the most conservative case. The host galaxy spectral continuum is best fit with a redshift of $z\sim2$, and other independent estimates converge to $z\sim1-2$. From the analysis of the early afterglow SED, we find that non-negligible host galaxy dust extinction, in addition to the Galactic one, affects the observed flux in the optical regime. The extinction-corrected optical-to-gamma-ray SED during the prompt emission shows a flux density ratio $F_{\gamma}/F_{opt}=10^{-2}-10^{-4}$ with spectral index $\beta_{\gamma,opt} > \beta_{\gamma}$, strongly suggesting a separate origin of the optical and gamma-ray components. This result is supported by the lack of correlated behavior in the prompt emission light curves observed in the two energy domains. The temporal properties of the prompt optical emission observed during \grb11 and their similarities to other rapidly-observed events favor interpretation of this optical light as radiation from the reverse shock. Observations are in good agreement with theoretical expectations for a thick shell limit in slow cooling regime. The expected peak flux is consistent with the observed one corrected for the host extinction, likely indicating that the starting time of the TAROT observations is very near to or coincident with the peak time. The estimated fireball initial Lorentz factor is $\Gamma\ge260 - 360$ at $z=1-2$, similar to the Lorentz factors obtained from  other GRBs. \grb11 is a rare case of a GRB with both a bright, well-observed optical counterpart and a `canonical' early X-ray light curve, thus providing a good test case of the reverse shock emission mechanism in both energy ranges. }
   {}
    }

\maketitle

\section{Introduction}

Measuring the multi-wavelength spectrum of the prompt emission of gamma-ray bursts (GRBs) is a challenge, requiring coordinated observations both from space and on the ground that are perfomed on very short notice (a few seconds). Such observations are nevertheless 
sorely needed to constrain the nature of the prompt emission. 
Rapid-response telescopes \citep[e.g.][]{Akerlof2003,Boer2001,Perez2004,Covino2004,Bloom2006} and developments in the technique of wide-field optical surveys \citep[e.g.][]{Vestrand2002,Pozanenko2004,Cwiok2005,Tamagawa2005}  are fundamental to this purpose.
So far, a few dozen GRBs have been detected 
at optical wavelengths when the GRB was still active, or soon after its end 
\citep[e.g.][]{Blake2005,Yuan2009,Gendre2008a}. In several cases (e.g. GRB 990123, GRB 080319B), 
the optical flux after the peak emission  
shows a very steep decay that renders the detection of this component even more challenging \citep[e.g.][]{Melandri2008}. 
The demanding rapidity of observing these bright optical flashes (about a dozen seconds after 
the trigger or less) is on average fulfilled by automatic ground-based telescopes but not by the onboard $UV$-Telescope {\it Swift}/UVOT, whose first observation times range from 
40 to 200 seconds after the trigger \citep{Roming2009}. 
It is thus difficult to infer the true fraction of GRBs showing this emission.

The common interpretation of the observed emission from GRBs relies on 
the fireball internal-external shock model (see e.g. \citeauthor{Zhang2007} 2007 for a recent review). 
Whether the origin of the optical counterpart observed during 
the prompt emission is from internal shocks, like the high-energy emission, 
or from an external (reverse) shock, is still an open debate in the context of this model \citep{Genet2007a}. 
Answering this question will provide useful insight into the 
formation of GRB shocks. In general, the early optical emission might be a superposition 
of internal shocks, reverse shock and forward shock, with different contributions 
at different phases of the time history.

The first Swift-BAT position notice of \grb11 was distributed at 20:16:03 UT, 20 s
after the trigger, allowing the TAROT and ROTSE III robotic telescopes to observe during prompt emission \citep{Klotz2006,Yost2006,Rykoff2009}. 
TAROT Calern (France) started observing at 20:16:11 UT (28 s after trigger) 
and detected a bright optical emission with R$\sim$13.7, fading as $t^{-2.4 \pm 0.1}$ until 80s 
\citep{Klotz2006}. 
The position of the afterglow as measured 
by MITSuME from 45-minute exposure images taken about 10 minutes after the trigger is R.A.= 19$^h$ 05$^m$ 42.47$^s$, Dec.= +70$^\circ$ $22'$ $33.1''$  (J2000.0), with a precision of $0.3''$ \citep{Yanagisawa2006}. This is fully compatible with the position measured by UVOT with an accuracy of $0.5''$ (RA = 19$^h$05$^m$42.48$^s$, Dec.= +70$^\circ$ $22'$ $33.6''$  \citeauthor{Perri2006a} 2006a). 

Because of the high declination of the source, ground-based optical follow-up 
has been challenging: the position cannot be observed 
from any major South Hemisphere observatory, and while it is circumpolar in 
the northern hemisphere, at late times the hour angle of the source is very large 
throughout the night and not accessible to equatorial telescope mounts.
Despite these unfavorable conditions we set-up an observing campaign
that provided many important results: 
the identification of the host galaxy, strong constraints
on any supernova associated with the GRB and 
on the host-galaxy redshift.
We present in Table \ref{tab-opt} the list of 
the observations of the GRB error box that 
led to the results presented in this paper.


The first part of this paper deals with the observations of \grb11 
(\S 2). Results from the multiwavelength prompt and afterglow analysis are 
presented in \S \ref{res1}. Results from the host galaxy identification, as well as 
the redshift constraints of the source, are presented in \S \ref{res2}, while in \S \ref{discuss-prompt} we discuss the nature of the prompt optical 
emission in light of our multiwavelength campaign results.

\section{Observations and data analysis}

\subsection{{\it Swift}/BAT and WAM/Suzaku}
\label{obs}

\begin{figure}[htbp]
\begin{center}
\includegraphics[width=6cm,angle=-90]{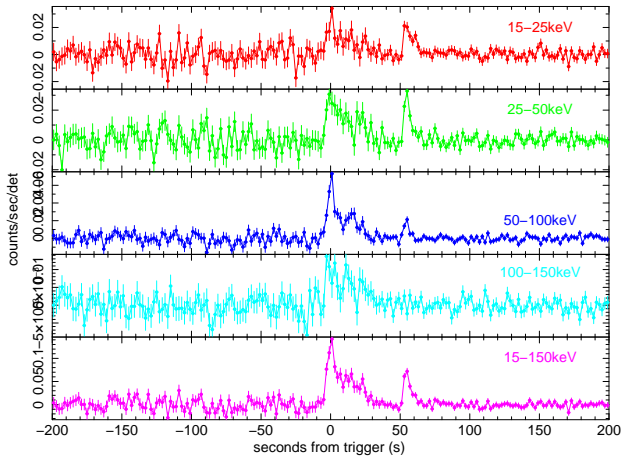}
\caption{Swift/BAT count rate of \grb11 in the ranges (from top to bottom): 15-25 keV, 25-50 keV, 50-100 keV, 100-150 keV and 15-150 keV. Count rate units are counts per seconds per illuminated-detector (note illuminated-detector = 0.16 cm$^2$).}
\label{multi}
\end{center}
\end{figure}

\grb11\ was a moderately bright GRB 
detected on 2006, January 11 at 20:15:43.24 UT (hereafter 
t$_{\mathrm{trig}}$) by the BAT instrument on the Swift spacecraft \citep{Perri2006a,Tueller2006a}.  
This burst exhibits two peaks reaching their 
maxima at t$_{\mathrm{trig}}$+2 s and t$_{\mathrm{trig}}$+55 s. While 
the first peak is clearly detected in the full BAT energy band, the second is 
significantly softer (Fig. \ref{multi}).
The duration of the burst is $T_{90}=59\pm5$ s in the range 15-150 keV. 
The time-averaged spectrum (from t$_{\mathrm{trig}}-2$ to t$_{\mathrm{trig}}+63$ s) is well-fit by a power-law model, with best-fit photon index $\Gamma_{\gamma}=0.9 \pm 0.1$. The observed peak energy flux and fluence in the 15-150 keV band are $2.6\times10^{-8}$ erg cm$^2$ s$^{-1}$ and $1.6 \times 10^{-6}$ erg cm$^{-2}$, respectively.

The prompt high-energy emission was also observed by WAM on Suzaku, with a 
duration of $T_{90}=25.4$ s and a fluence of $5.6 \pm 0.8 \times 10^{-6}$ erg/cm$^{2}$ in the 100 - 1000 keV band \citep{Sato2006}.
A joint spectral analysis of the prompt emission with Swift/BAT and 
Suzaku/WAM data assuming a Band model, provides a best-fit, observer-frame peak energy of 
\ep\ = $462^{+290}_{-242}$ keV \citep{Fukazawa2007}.
This value is high enough to consider the spectral index measured by Swift
to be the low-energy spectral index of the Band function describing the 
energy spectrum of the prompt emission.

\begin{figure}[htbp]
\begin{center}
\includegraphics[width=\columnwidth]{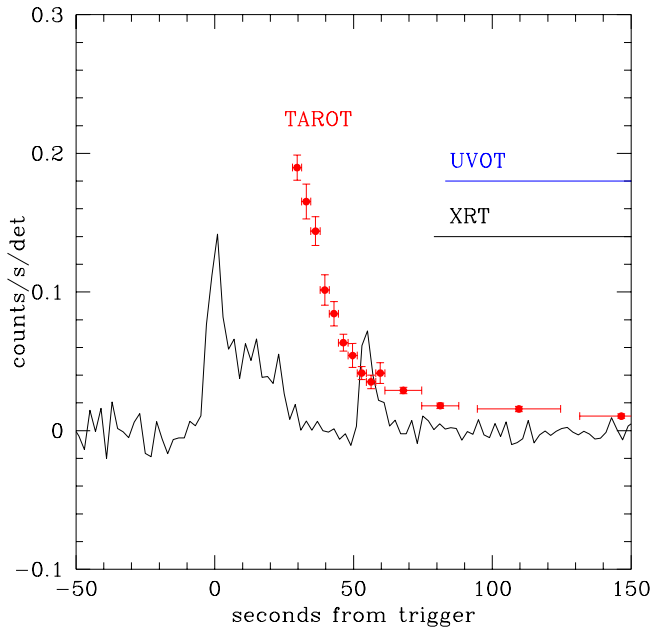}
\caption{Swift/BAT count rate (in units of counts per seconds per illuminated-detector) of \grb11 in the range 15-150 keV. The TAROT light curve  superimposed on the BAT count rate is $c_R=k10^{-0.4m_R}$ with $k=10^4$ in arbitrary units. The temporal ranges covered by UVOT and XRT data are also plotted.}
\label{fig:batlc}
\end{center}
\end{figure}

\subsection{{\it Swift}/XRT}
\label{obs-aft-xray}

{\it Swift} slewed immediately to the burst, and the X-Ray Telescope (XRT, \citeauthor{Burrows2005} 2005) began data acquisition at t$_{\mathrm{trig}}+87$ s in Windowed Timing (WT) mode. At t$_{\mathrm{trig}}+153$ s the count rate was at a critical level for XRT acquisition mode and an automatic flickering between the WT mode and the Photon counting (PC) mode started. At that time, the afterglow was too bright for the PC mode, and the PC data were affected by moderate pile-up.  At t$_{\mathrm{trig}}+426$ s, the XRT switched permanently to the PC mode and continued observations until t$_{\mathrm{trig}}+2.3$ days. In the following, we used WT data taken between t$_{\mathrm{trig}} + 87$ s and t$_{\mathrm{trig}} + 426$ s, and PC data taken later on. 
The XRT data were processed following standard procedures (\citeauthor{Capalbi2005}  2005\footnote{http://swift.gsfc.nasa.gov/docs/swift/analysis/}), by using the version 0.11.6 of the dedicated pipeline {\rm xrtpipeline}. Grade filtering was applied by selecting the 0-2 and 0-12 ranges for the WT mode and PC mode data, respectively. During the first orbit, the PC mode data show a count rate above 0.5 counts s$^{-1}$ thus likely affected by moderate pile-up. From point-spread function fitting \citep{Moretti2005} we excluded a circular region of 1 px radius from the extraction region of the source. 

The X-ray (0.3-10 keV) light curve from 87 s  to 2.3 days after the trigger shows a steep initial decay up to 150 s after the trigger that is poorly approximated by a power law. Starting from 150 s after the trigger we find that the data are best fitted by a double broken power law model, with decay indexes  $\alpha_{X,1}=1.25\pm0.15$ up to 1 ks, $\alpha_{X,2}=0.6\pm0.2$ up to $t_b=6\pm1$ ks and $\alpha_{X,3}=1.4\pm0.1$ later on ($\chi^2/d.o.f.=40/43$). These values are quite typical for Swift/XRT afterglow light curves \citep[e.g.][]{Liang2007}.  We note that these results are marginally consistent with the results from an automatic analysis of 318 GRBs by \cite{Evans2008} where, for \grb11,  $\alpha_{X,2}\sim1$ from 175 s to $10^4$ s after the trigger, corresponding to the average of our $\alpha_{X,1}$ and $\alpha_{X,2}$. By fitting PC data only, a simple power law and a broken power law model are rejected with $\ge99.99\%$ and $\sim80\%$ confidence levels, respectively.

\begin{figure}
  \centering
   \includegraphics[width=8cm]{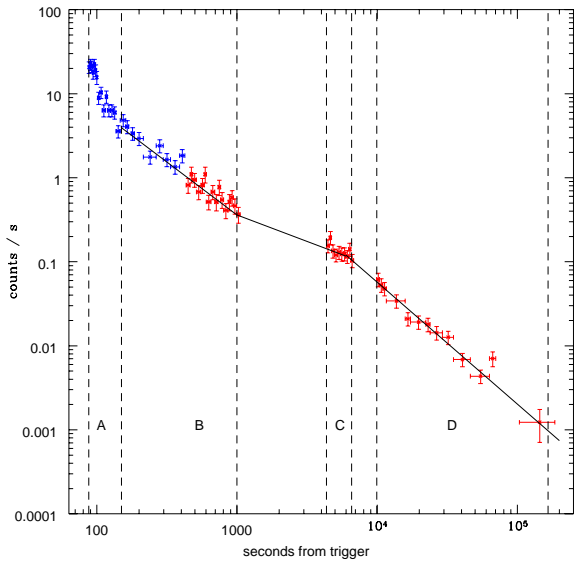}
   \caption{The 0.3-10 keV light curve (blue data are in taken in Windowed Timing mode and red data are taken in Photon Counting mode by XRT). The continuum solid line shows the best fit temporal model (see \S 2.2). For each plotted temporal interval (A,B,C and D) we extracted an energy spectrum (Tab. \ref{xaft}).   }
              \label{xlc}%
    \end{figure}

In order to investigate the spectral properties of each segment individuated in the light curve, we extracted the 0.3-10 keV spectrum of the X-ray afterglow at 4 different epochs, indicated with letters A, B, C, and D as plotted in Figure \ref{xlc}. Assuming an absorbed power-law model we found no evidence of spectral variation during the evolution of the afterglow, with a photon index of $\sim2$ and an equivalent hydrogen column density of $N_H\sim 2\times10^{21}$ cm$^{-2}$ in addition to the Galactic one ($N_{H,Gal}=7.65\times10^{20}$ cm$^{-2}$ from \citeauthor{Kalberla2005} 2005). These results are summarized in Table \ref{xaft}. Our temporal and spectral results are in agreement with results found by \cite{Racusin2008a}.

\begin{table*}[ht]
\label{tab-uvot}
\caption{Swift/UVOT single exposure frames analyzed following standard procedure and photometry results.} 
\begin{center}
\begin{tabular}{lccccccc}
\hline
\multicolumn{1}{c}{DATE } & \multicolumn{1}{c}{Filter}& \multicolumn{1}{c}{Exposure}  & \multicolumn{1}{c}{mag$^{\mathrm{a}}$} & \multicolumn{1}{c}{mag$^{\mathrm{a}}$} & \multicolumn{1}{c}{$F_{\nu}^{\mathrm{a}}$} & \multicolumn{1}{c}{Log frequency} &  \multicolumn{1}{c}{Log ($\nu$ $F_{\nu}$ )} \\ 
\multicolumn{1}{c}{UT } & & (s) & &(dereddened) & ($10^{-28}$erg cm$^{-2}$s$^{-1}$Hz$^{-1}$) &  &    \\ 
\hline
2006-01-11 20:17:06   &  $V$    &   196.6 & 17.59$\pm$0.11 & $17.24\pm0.11$  &   46.19 $\pm$ 4.67  & 14.74  &  -11.59 $\pm$0.10  \\
2006-01-11 20:29:35   &  $V$    &   196.6 & $>$ 18.86      & $>$18.51        & $>$ 14.34               & 14.74  &  -12.10 $\pm$0.00  \\
2006-01-11 21:52:36   &  $V$    &   118.8 & $>$ 18.86      & $>$18.51  &  $>$ 14.34               & 14.74  &  -12.10 $\pm$0.00  \\
2006-01-11 23:12:09   &  $V$    &   196.6 & $>$ 19.29      & $>$18.94  &  $>$ 9.65               & 14.74 &  -12.27 $\pm$0.00  \\
2006-01-11 20:20:33   & $B$   &   196.6 & $19.73\pm0.32$ & $19.27\pm0.32$  &   79.46 $\pm$ 2.34  & 14.84  &  -12.26 $\pm$0.29  \\
2006-01-11 20:33:01   & $B$   &   93.4 & $>$ 19.24      & $>$  18.78  &  $>$ 12.48               & 14.84  &  -12.06 $\pm$0.00  \\
2006-01-11 22:06:15   & $B$   &   65.5 & $>$ 18.98      & $>$  18.52  &  $>$ 15.85               & 14.84  &  -11.96 $\pm$0.00  \\
2006-01-11 23:37:36   & $B$   &   221.2 & $>$ 19.84      & $>$   19.38  &  $>$ 71.80               & 14.84  &  -12.30 $\pm$0.00  \\
2006-01-11 23:41:24   & $B$   &   221.2 & $>$ 19.67      & $>$   19.21  &  $>$ 83.97               & 14.84  &  -12.23 $\pm$0.00  \\
2006-01-11 23:45:12   & $B$   &   126.6 & $>$ 19.43      & $>$   18.97  &  $>$ 10.47               & 14.84  &  -12.14 $\pm$0.00  \\
\hline
\end{tabular}
\begin{list}
\item[$^{\mathrm{a}}$] The upper limits are given at 3 $\sigma$ confidence level.
\end{list}
\end{center}
\end{table*}

\subsection{{\it Swift}/UVOT}
\label{obs-uvot}

The Swift follow-up with UVOT started 84 s after the trigger
\citep{Perri2006a,Perri2006b,Holland2006}. 
We analyzed the UVOT images starting with the reduction of raw 
data using the standard Swift\/UVOT software 
reduction tasks to create the sky images.
We performed the analysis 
with the release 2.7.2 (HEADAS version 6.3.1) 
and the 20070711 version of the CALDB.
 
Our analysis involved the steps listed below. 
We performed the ``standard'' aperture photometry of the object 
in the sky images for each single exposure and, as comparison, 
in the images obtained as the sum of all the exposures in each filter. 
Photometry was performed using the UVOT-HEADAS tasks ``uvotsource'' and ``uvotmaghist''.
For this purpose we first selected the best source and background aperture regions 
from the first single exposure frame of the $V$ filter. We applied the standard sizes for the source and background apertures, excluding the nearby contaminating objects from the background extraction apertures. The results of the photometry of each single exposure in each filter is reported in Table 1. We also report the dereddened flux obtained by correcting the observed magnitude using a mean Galactic interstellar extinction law \citep{Fitzpatrick1999} with a value of $E(B-V)=0.11$ mag \citep{Schlegel1998}.  
The logarithm of the $\nu F_{\nu}$ calculated from this flux, i.e. the Spectral Energy Distribution, is also reported.

\begin{figure}
   \centering
   \includegraphics[width=6cm]{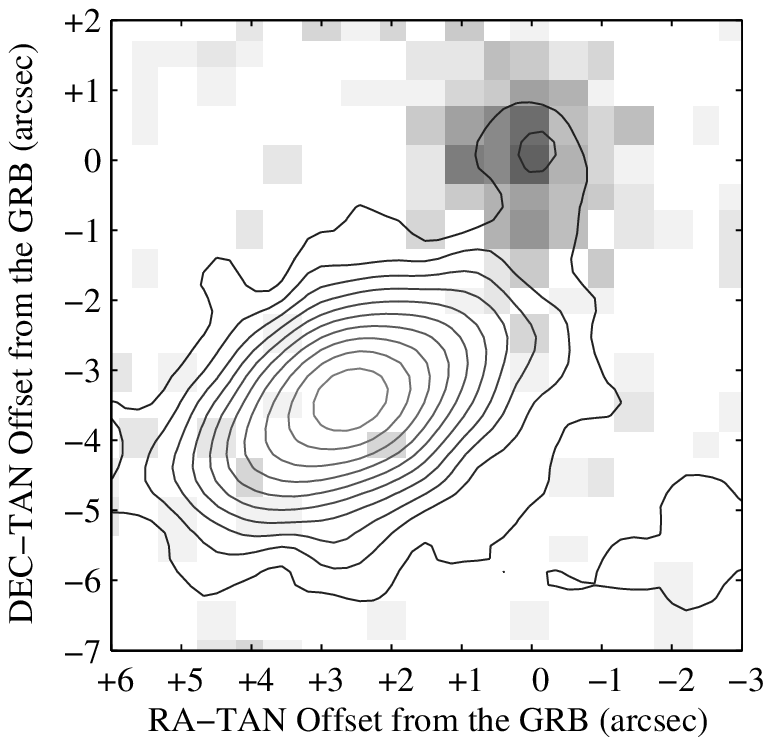}
   \includegraphics[width=6cm]{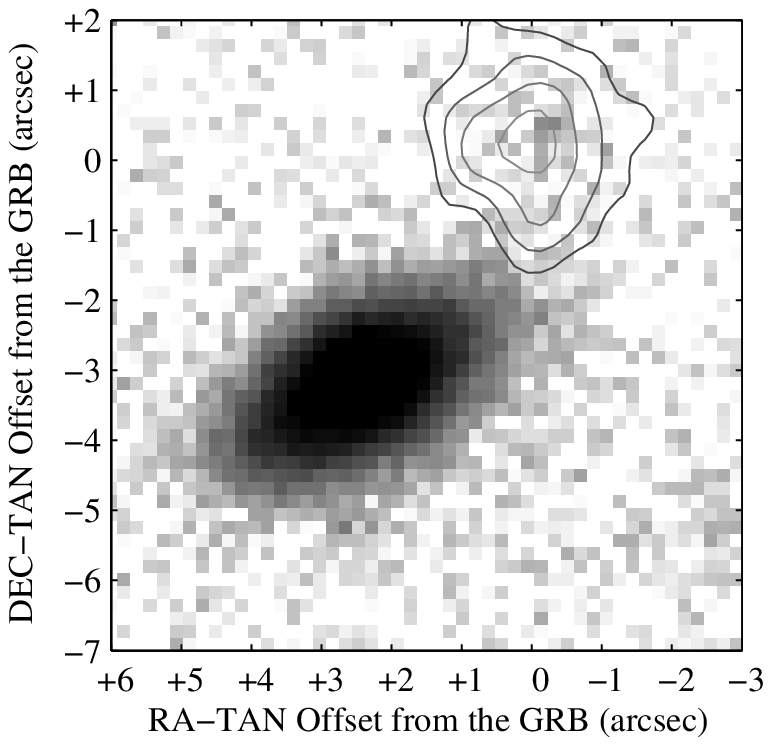}
      \caption{{\it Top panel}:  $R$ -band isophote curves of the host galaxy and the nearby bright galaxy 
from the CFHT image taken 159 days after the trigger. 
Superposition of the UVOT image taken 123 s after the trigger shows that the afterglow 
coincides with the host galaxy candidate from the CFHT image.  {\it Bottom panel}: CFHT gray scale 
image is superposed to the UVOT isophotes of the afterglow.
    }
         \label{fig:f9}
   \end{figure}

We further refined the analysis of the first single exposure frame of the UVOT image of filter  $V$  by applying  
 ``non-standard'' apertures for the source extraction and by applying the aperture 
correction procedure in order to enhance the signal to noise ratio.
Various ``non standard'' extraction regions, such as with larger radii or different shapes 
(for example an elliptical region), have been tested, but generally give results similar to those 
obtained using the standard extraction radius. 
For the aperture correction procedure a first photometry on apertures of 1.9$''$ 
(the FWHM of the  $V$  filter) was performed.
Then 6 objects not saturated in the field were selected and photometry 
was performed using FWHM apertures and standard apertures (5$''$ radius) for each of them.
The difference of the magnitudes in the two different apertures was evaluated 
for the stars selected and finally the weighted mean among all the differences was calculated.
This value was subtracted to the GRB magnitude obtained with FWHM aperture.
The final value of the GRB magnitude thus obtained integrating over all the exposure 
time is $V=17.55\pm 0.08$ mag that is consistent with our standard analysis. 
The upper panel of Figure \ref{fig:f9} shows the field of the GRB in
filter V, during the first single exposure frame of sky image.

In order to increase the temporal resolution, 
the event data in all filters were also analyzed to extract the
UVOT lightcurve of the object. 
Event data were first screened to filter out unwanted
contaminations using standard selection criteria, as excluding the
South-Atlantic Geomagnetic Anomaly and possible anormal attitude
variations, keeping the angle between the pointing and the Earth limb
greater than 10 degrees and the same angle with ``bright'' Earth limb
greater than 20 degrees. Moreover, only events classified as ``good''
were accepted. The lightcurve for each filter was 
extracted using the UVOT task ``uvotevtlc'', with different binning for
the  $V$  filter (30 {\rm s} bin) as compared to the remaining (70 {\rm s}
bin except for the White for which 100 {\rm s} bin has been used).
The photometry is calibrated using the photometric zero points in 
the Swift/UVOT calibration data base. These zero point have a systematic 
error of 0.1 mag that is not included in the values quoted in Table 2.

   \begin{table*}
      \caption{Optical observations of \grb11 not corrected for Galactic extinction. }
         \label{tab-opt}
     $$
         \begin{tabular}{rrrrrl}
            \hline
            \noalign{\smallskip}
T$_\mathrm{mid}^{\mathrm{a}}$ & $T_{exp}$ (s) & Filter & mag$^{\mathrm{b}}$ & $\Delta$mag$^{\mathrm{b}}$ & ref. \cr
            \noalign{\smallskip}
            \hline
            \noalign{\smallskip}
      29.65  s &      3.3    &  $R$ & 13.75    &    0.06  & TAROT$^{\mathrm{(1)}}$ \cr
       32.95 s   &     3.3   &  $R$ &   13.9    &    0.09 & TAROT \cr
        36.3 s  &      3.4   &  $R$ &  14.05    &    0.09 & TAROT \cr
       39.65 s  &      3.3   &  $R$ &  14.43   &     0.13 & TAROT \cr
       42.95 s  &      3.3   &  $R$ &  14.63    &    0.13 & TAROT \cr
        46.3  s &      3.4   &  $R$ &  14.94   &     0.12 & TAROT \cr
       49.65 s  &      3.3   &  $R$ &  15.11   &      0.2 & TAROT \cr
       52.95 s  &      3.3   &  $R$ &   15.4   &     0.14 & TAROT \cr
        56.3 s  &      3.4   &  $R$ &  15.58  &      0.17 & TAROT \cr
       59.65 s  &      3.3   &  $R$ &   15.4   &     0.22 & TAROT \cr
       67.95 s &      13.3   &  $R$ &  15.79  &      0.10 & TAROT \cr
       81.25 s  &     13.3   &  $R$ &  16.31  &      0.13 & TAROT \cr
       109.6 s  &        30   &  $R$ &  16.46  &       0.15 & TAROT \cr
       146.5 s &        30   &  $R$ &  16.89  &      0.20 & TAROT \cr
       183.3 s  &       30   &  $R$ &  17.53  &      0.30 & TAROT \cr
85.5  s& 30  & $V$ & 16.8    & 0.4 & UVOT     \cr
115.5 s& 30  & $V$ & 17.2    & 0.5 & UVOT     \cr
145.5 s& 30  & $V$ & 17.7    & 0.7 & UVOT     \cr
175.5 s& 30  & $V$ & 17.6    & 0.7 & UVOT     \cr
388 s  & 197 &$B$& 19.73  & 0.32 & UVOT     \cr
792  s  & 300   & $R$ & 18.9     & 0.4  & MITSuME$^{\mathrm{(2)}}$  \cr
792   s & 300   & $I$ & 18.3     & 0.4  & MITSuME$^{\mathrm{(2)}}$  \cr
0.28 days&  3900   & $R$ & 23.2  &   0.5   &   RTT-150       \cr
12.19 days & 1440  & $R$ & $>$21.7  &      &   Maidanak-1.5  \cr
24.18 days  & 2700 & $R$ & $>$22.3  &      &   Maidanak-1.5  \cr
27.20 days  & 1800 & $R$ & $>$22.7  &      &   Maidanak-1.5  \cr
31.18 days & 3420  & $R$ & $>$22.5 &       &   Maidanak-1.5  \cr
16.29 days & 3360  & $R$ & 25.1     & 0.5  &   ZTSh     \cr
108.0  days    &  1200  	 & $V$ & $>$21.6  &      & ZTSh \cr
26.3 days  & 3600  & $R$ & $>$21.0  &      & OHP-T120 \cr
37.1 days & 2700  & $R$ & $>$22.5 &       &   NOT           \cr
159 days     &  1840 	 & $R$ & 24.8     & 0.4  & CFHT     \cr
2.3 years     &   1800	 & $H$ & $>23.0$     &   & CFHT     \cr
\hline
\hline
5.3 months      &  4800  & &3800-8500\AA & spectroscopy   & Keck/LRIS     \cr
            \noalign{\smallskip}
            \hline
         \end{tabular}
     $$
\begin{list}{}{}
\item[$^{\mathrm{a}}$] $T_{mid}$ is the mean epoch of the exposure of duration $T_{exp}$
\item[$^{\mathrm{b}}$] Errors are given at 1$\sigma$. The upper limits are given at 3 $\sigma$ confidence level. 
\end{list} 
References. (1) all TAROT data are from \cite{Klotz2006}; (2) \cite{Yanagisawa2006}
   \end{table*}

\subsection{Ground based optical follow-up}
\label{obs-aft}

Very early optical emission, measured with unprecedented temporal resolution of a few seconds by the TAROT robotic observatory, has been presented by \cite{Klotz2006}. 
We quote here their results in Table \ref{tab-opt}.   
ROTSE-IIId observations that started 33 s after the trigger and detected a 
bright optical flare which was initially at $\sim13$ mag and 
fading as $\sim t^{-(2.35\pm0.10)}$, in agreement with TAROT observations 
\citep{Yost2006,Rykoff2009}. The field of GRB 060111B was also observed with the three-color MITSuME 50 cm telescope at Okayama, Japan starting at $t_{trig}+10.6$ minutes in the Rc and Ic band images \citep{Yanagisawa2006}.

Few hours after the trigger, \grb11 was observed with the Russian-Turkish 1.5-m telescope (RTT150, Bakyrlytepe, TUBITAK National Observatory, Turkey). 
In order to extract the maximum information on the OT, we carefully analyzed 
the sum and median stacks of 13 images of 300s taken at RTT150 between 5.92 and 7.20 
hours after the trigger (see also \citeauthor{Khamitov2006} 2006).
The afterglow is marginally detected with a signal to noise ratio of S/N=2.
A PSF fit of the theoretical location of the afterglow gives $R=23.2 \pm 0.5$.

Further observations were performed with AZT-22 (Maidanak Astronomical 
Observatory, Uzbekistan) 2.6m Shajn reflector (ZTSh, Crimean Astrophysical Observatory, Ukraina), OHP T120 (Observatoire de Haute-Provence, France) 
and NOT with only one detection at ZTSh 16.3 days after the trigger (see Table \ref{tab-opt}). 

Late time observations of the host galaxy are discussed in section \ref{sub-imaging}.

   \begin{figure}
   \centering
   \includegraphics[width=9cm]{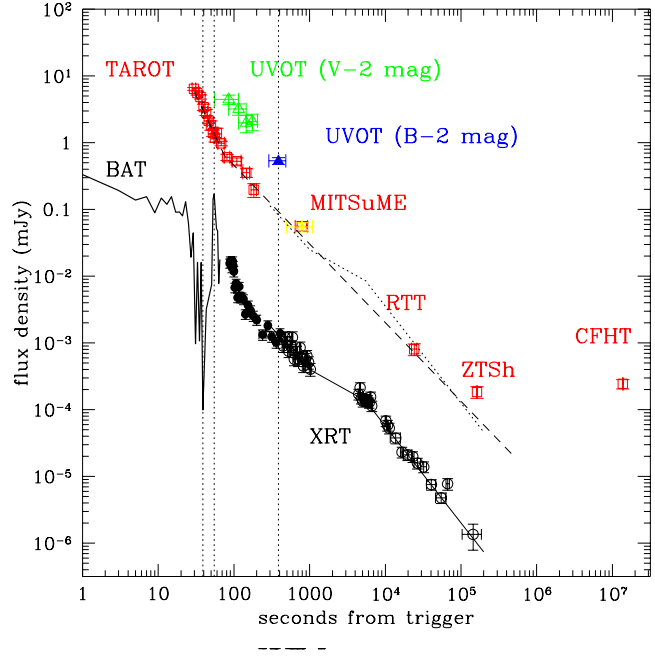}
      \caption{Temporal evolution of the prompt and afterglow flux at several wavelengths. The  $R$ -band early data are clearly not correlated with the BAT prompt emission. The dashed curve is the  $R$ -band data best fit model while the solid curve is the X-rays best fit model. The dotted curve is the best fit X-ray model plotted over the  $R$ -band data. The three dotted vertical lines are the epochs at which a SED was extracted. 
               }
         \label{fig:optlc}
   \end{figure}

\section{Multiwavelength analysis}
\label{res1}

\subsection{Temporal evolution}

The prompt emission of \grb11 was observed simultaneously in 
the optical band ( $R$ ) and in the gamma-rays (15-150 keV), starting from 28 s 
up to 60 s ($T_{90}$) after the trigger time. During this interval, the optical emission was found to decay steadily, in contrast with the gamma-ray emission 
which exhibited flux variability of 2 orders of magnitude \citep{Klotz2006}. 

The prompt optical emission from $t_{trig} +28$ s to $t_{trig} +200$ s does not track the X-ray (0.3-10 keV) emission either. The X-ray light curve, from $\sim t_{trig} + 87$ s to $\sim t_{trig} + 150$ s, shows a much steeper decay. However, starting from 200 s after the trigger the X-ray and optical decay indexes are consistent (see Tab. \ref{xaft} and Fig. \ref{fig:optlc}). Indeed, the TAROT data, the MITSuME   $R$ -band observations (taken 13 minutes after the trigger, \citeauthor{Yanagisawa2006} 2006) and the RTT150 observations, show that 
the  $R$ -band light curve from 28 s to 7.2 hrs after the trigger can be modeled with a broken power law with 
best fit decay indexes $\alpha_{R_1} = 2.4 \pm 0.2$ and $\alpha_{R_2} = 1.18 \pm 0.05$ with a break at $\sim75$ s. The  $V$ -band decay index between $\sim t_{trig} + 85$ and $\sim t_{trig} +200$ s is $\alpha_V = 1.2 \pm 0.4$, consistent with the  $R$-band decay index. The ZTSh detection (taken 16.3 days after the trigger) slightly exceeds the power-law extrapolation, likely due to the superposition of the underlying host galaxy (see \S 4.1). 
 
GRB 060111B represents one of the few cases where color 
information during the early afterglow is available. The similarity 
of the $V$ and $R$ band decay indexes (within uncertainties) between 85 s and 200 s after 
the trigger is indicative of achromatic evolution of the late afterglow. 
However the large uncertainty affecting the $V-$band decay index $\alpha_V$ 
and the proximity of $R$ and $V$ pass-bands prevent from more detailed investigation. The lack of a continuous coverage in the optical band after about $t_{trig}+$1 ks, makes possible a consistency of the  $R$ -band light curve at late times with the X-ray shallow-to-normal behavior with a temporal break at about 6 ks after the trigger.

\subsection{Spectral Energy Distribution}

\begin{figure}[htb]
\sidecaption
   \includegraphics[width=7cm]{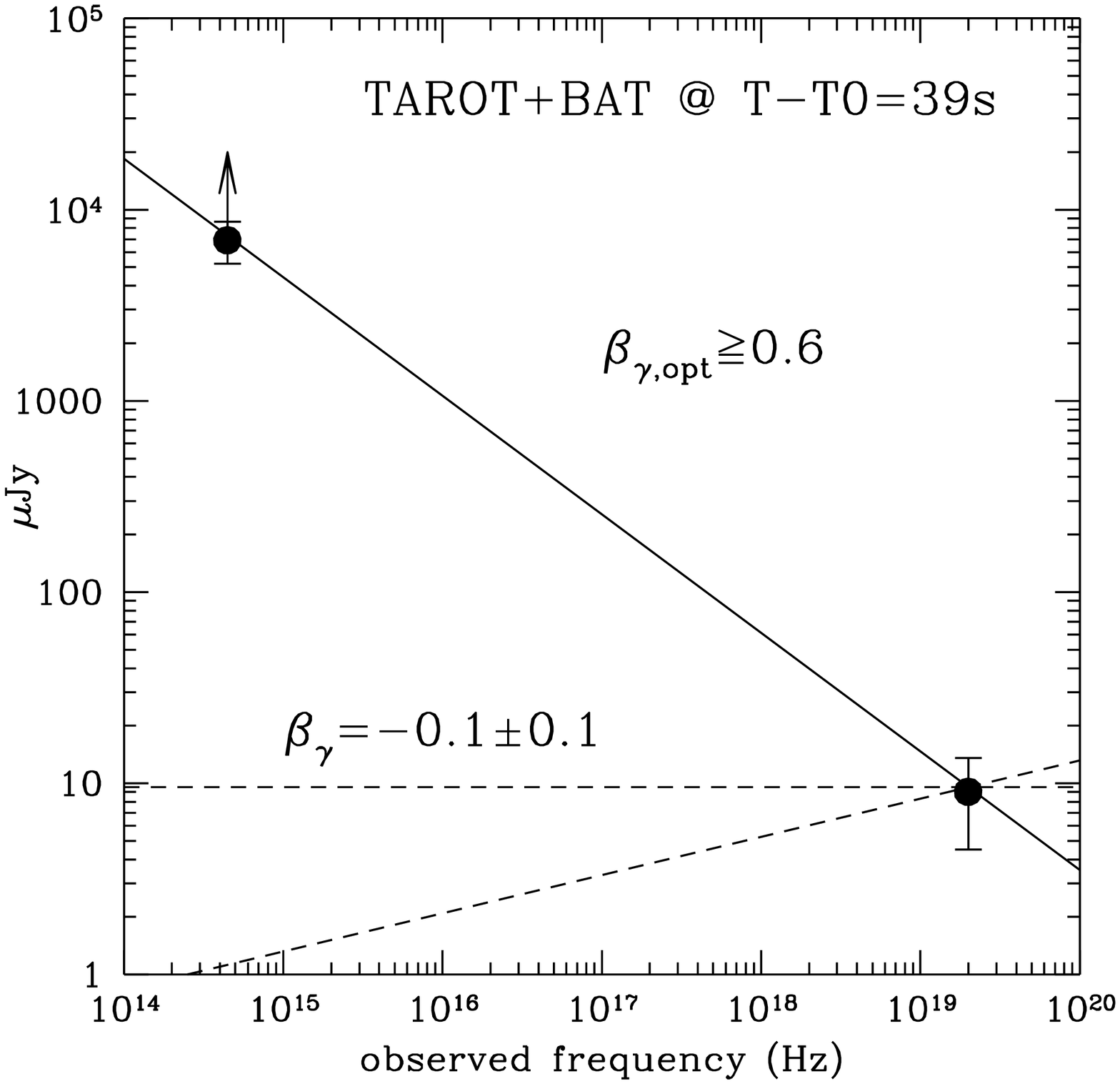}
   \includegraphics[width=7cm]{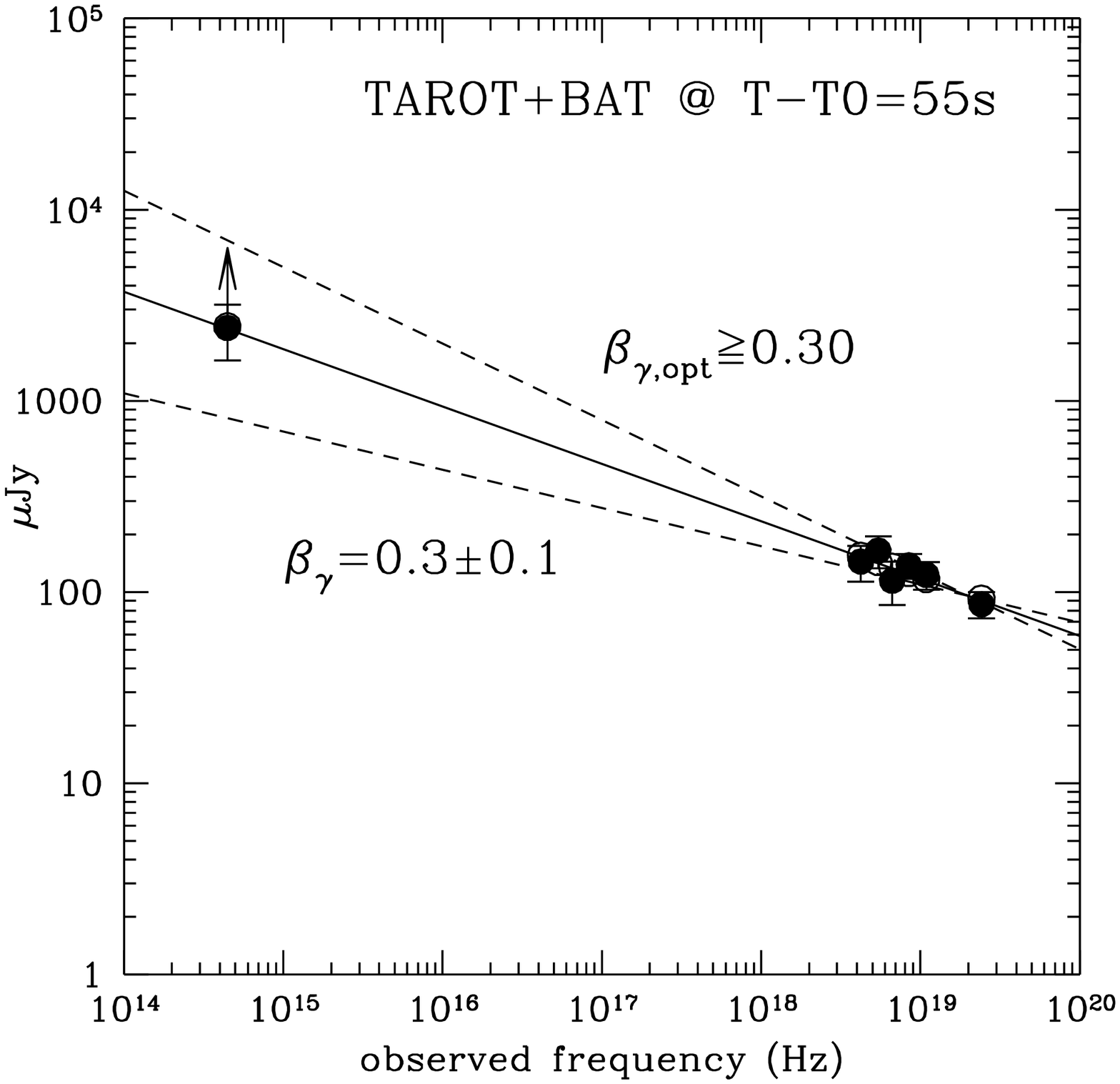}
\caption{
Spectral Energy Distribution at $t_{trig}+39$ s and $t_{trig}+55$ s after the trigger where optical data have been corrected for Galactic extinction \citep{Schlegel1998} but not for host galaxy extinction and should be considered as lower limits. The solid curve represents the optical-to-gamma-ray slope while the dashed lines individuate the best fit power law model from gamma-ray data only within 1 $\sigma$ (at $t_{trig}+39$ s we have assumed the $T_{90}$ integrated spectrum best fit model). 
} 
\label{fig:SED2}
\end{figure}

In order to investigate on the origin of the prompt optical 
emission, we analyzed the Spectral Energy Distribution (SED) 
from simultaneous BAT and TAROT observations. 
Two SEDs were extracted at 39 s after 
the trigger (from 28 s to 50 s, immediately before the second high energy prompt-emission peak) and at 55 s (from 50 s to 60 s, during the second peak).
In the first interval, the available statistics from BAT data is too low to obtain spectral information and we could compute only the observed average 15-150 keV count rate, that is $1.6\times10^{-3}$ counts s$^{-1}$. We converted this count rate into flux density assuming the $T_{90}-$averaged best fit spectral model (a simple power law with a photon index of $\Gamma_{\gamma}=\beta_{\gamma}+1=0.9\pm0.1$ see \S 2.1). Extrapolating the power law model back to low energies, the TAROT flux density is about 3 orders of magnitude above the extrapolation of the gamma-ray flux density, with $F_{\gamma}/F_{opt}\sim 10^{-3}$ and $\beta_{\gamma,opt}\sim0.6$ (Fig. \ref{fig:SED2}). 
During the second peak of the burst (i.e. 55 s after the trigger), the available statistics enabled to extract a high energy spectrum best fitted by a simple power law with photon index of $\Gamma_{\gamma}=\beta_{\gamma}+1=1.3\pm0.1$. Fixing this value within the uncertainties, we found that the  $R$ -band extrapolation is consistent with the observed flux (Fig. \ref{fig:SED2}). The $F_{\gamma}/F_{opt}$ is now $\sim0.1$ and $\beta_{\gamma,opt}\sim\beta_{\gamma}=0.3$. 

These results assume no additional extinction beyond that corrected for in our Galaxy. 
However, from X-ray spectral analysis there are hints of non-null absorption from the host galaxy. For example, assuming 
a Milky Way ISM, the rest frame $A_V^{\rm host}$ is expected to be $N_H/1.8\times10^{21}$ mag cm$^{-22}$ \citep{Predehl1995}, that for \grb11 is of about 1 mag or more since our $N_H$ measure ($N_H\sim2\times10^{22}$ cm$^{22}$ assuming $z=0$, \S 2.2) should be considered as a lower limit given the unknown redshift. We thus  attempted to estimate the host galaxy dust extinction affecting the TAROT data by exploiting our multiwavelength data set. We used Swift/XRT 0.3-10 keV and Swift/UVOT $B$-band and   $V-$band data as well as the TAROT  $R-$band and MITSuME $I-$band data. Since we have only one detection in the $B-$filter, the most sensitive wavelength to dust extinction,  we decided to extract the SED at the time of the $B-$filter detection, that is at 388 s after the trigger.
$I-$, $R-$ and $V-$band data have been extrapolated at $t_{trig}+$388 s assuming the $R$-band decay slope at that time, that is $1.18\pm0.05$, and then corrected for Galactic extinction, that in each filter is  
A$_B$=0.48, 
A$_V$=0.37,
A$_R$=0.30 and
A$_I$=0.22 mag, using a standard galactic extinction curve.  


The optical spectral index is unusually steep ($\beta_{opt}\ge2$). For comparison, other optical spectral indexes of GRBs with early optical steep-to-shallow light curve for which spectral data are available during the shallow phase are e.g. $\beta_{opt} = 0.17$ for GRB 080319B \citep{Wozniak2009} and $\beta_{opt} = 0.95$ GRB 061126 \citep{Perley2008}. This lends additional support to our inference of significant extinction in the host galaxy.

The consistency of the decay slopes in optical and X-rays at $t_{trig}+$388 s suggests a common cooling regime of the electron populations responsible for the observed emission in the two energy ranges \citep{Sari1998}. Specifically, the flux decay rate and the X-ray spectral index are consistent with the fireball expected closure relationship for an electron population in slow cooling regime and emitting at frequencies above the cooling frequency $\nu>\nu_c$. The expected spectral and decay indexes are $p/2$ and $(2-3p)/4$ respectively, where $p$ is the electron distribution spectral index. We find a good agreement with the measured indexes for $p\sim2.3$.  The optical and X-ray SEDs, both extracted at $t_{trig}+$388 s, were therefore fitted simultaneously assuming a simple power law model with spectral index fixed (within 1 $\sigma$) at the value obtained from X-ray data analysis only, that is $\alpha_X=1.2\pm0.1$ (e.g.  interval B in Table \ref{xaft}). To account for dust extinction from the host galaxy, we tested three different extinction curves.  

\begin{figure}[htb]
\sidecaption
	\includegraphics[width=7cm]{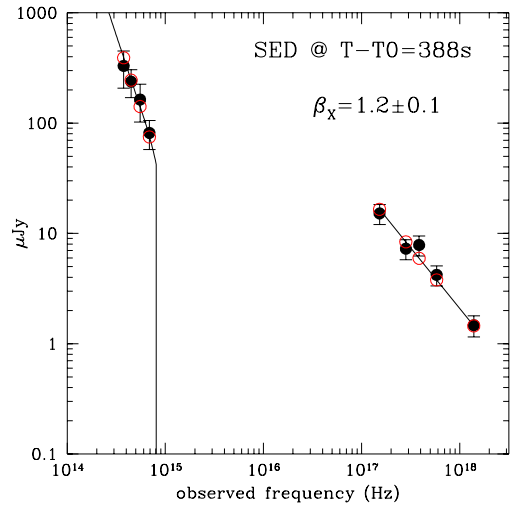}
\caption{
Spectral Energy Distribution at $t_{trig}+388$s where optical data have been corrected for Galactic extinction \citep{Schlegel1998}.  The solid curve is the optical-to-X-ray best fit model obtained at $z=2$ assuming a Starburst extinction curve (see \S 3.2) and the open red circles are the expected flux density values in each energy band. 
} 
\label{fig:SED1}
\end{figure}

Assuming different redshifts, the best fit among Galactic \citep{Cardelli1989}, SMC \citep{Pei1992} and Starburst \citep{Calzetti1994} extinction curves was obtained with the latter at $z=2$ with rest frame visual extinction $A_V^r\sim2$ mag (Fig. \ref{fig:SED1}, Tab. \ref{sed}). We have assumed a ratio of total to selective extinction of $R_V=A_V/E(B-V)=3.1$ for the Galactic and the Starburst extinction curves while $R_V=2.93$ for the SMC extinction curve \citep{Pei1992}. At $z<2$ acceptable fits  were obtained only with the Starburst extinction curve while at $z\ge2$ comparable good fits were obtained also assuming the Galactic extinction curve up to $z=2.5-2.7$, above which the Lyman $\alpha$ cut-off due to inter-galactic neutral hydrogen absorption would provide a $B-$band flux density more than 2$\sigma$ below the observed one.

At $z=2$, the observed  $R-$band corresponds to the $UV$ band in the rest frame (central wavelength $\sim 0.25 \mu$m), with best fit rest frame extinction of $A_{UV}=2.5\pm0.7$ mag. We note that at $z=1$ and $z=3$ the corresponding rest frame $UV$ extinction is 2.7 and 1.7 mag, respectively. Indeed, if at $z\ge2$, part of the $B$-band attenuation may be due to Lyman $\alpha$ cut-off and not only to dust extinction. Despite the redshift uncertainties, these findings indicate that the observed optical flux is attenuated by host dust extinction.   
Correcting the observed  $R-$band magnitude of $\sim2.5$ mag, that is the best fit value with a starburst extinction curve, the optical flux density increases by about a factor of 10. The above analysis of the gamma-to-optical SED would now provide $F_{\gamma}/F_{opt}\sim 10^{-4}$ and $\beta_{\gamma,opt}=0.80\pm0.05$ at 39 s after the trigger and $F_{\gamma}/F_{opt}\sim0.01$ and $\beta_{\gamma,opt}=0.50\pm0.05$ at 55 s after the trigger.

These results make \grb11 consistent with the few GRBs with prompt optical detection for which the extrapolation of the prompt gamma-ray flux to the optical bands assuming a simple power law under-predicts the observed flux (cases `B' and `D' in \citeauthor{Yost2007} 2007). Together with their different evolution in time, these findings suggest two separate origins for the prompt gamma-ray and the optical emission components. 

In \cite{Klotz2006} a value of $A_V^r\sim4$ mag at $z\sim1$ was reported from our 
XRT and TAROT data at $\sim t_{trig}+140$ (assuming a simple underlying power-law spectral continuum and a Galactic extinction curve). This is significantly higher than our estimate at $t_{trig}+$388s, possibly indicating an active dust destruction mechanism or the presence of a spectral break between the two bands at that time. In the latter case, the decaying flux behavior and the similarity of the decay rates observed at optical wavelengths and in X-rays from $\sim200$ s after the trigger suggest that the synchrotron cooling frequency crossed the rest frame UV band between $t_{trig}+140$ s and $t_{trig}+200$ s after the trigger.

\begin{table}
\centering
\caption{Best fit parameters obtained for the Spectral Energy Distribution extracted at $t_{trigg}+388$s assuming a simple power law model. }
\begin{tabular}{cccc}
\hline
Redshift & Extinction Model  & $A_V^r$ & $\chi^2/\nu$  \\
 	 & 	  & mag     &               \\
\hline
0	 & Galactic & $5.2\pm0.2$ & 19.8/7 \\
	 & SMC      & $5.5\pm0.2$ & 20.7/7 \\
         & Starburst& $4.1\pm0.3$ & 5.2/7 \\
\hline
1	 & Galactic & $2.2\pm0.2$ & 14.0/7 \\
	 & SMC      & $2.1\pm0.2$ & 15.2/7 \\
         & Starburst& $2.8\pm0.2$ & 3.5/7 \\
\hline
2	 & Galactic & $1.5\pm0.1$ & 5.1/7 \\
	 & SMC      & $2.1\pm0.1$& 15.2/7 \\
         & Starburst& $2.3\pm0.2$& 2.8/7 \\
\hline
\hline
\end{tabular}
\label{sed}
\end{table}

\begin{table*}
\centering
\caption{Spectral and temporal parameters at different epochs and energy ranges. }
\begin{tabular}{llclllc}
\hline
Epochs           & Instrument & Energy range & Decay Index & Spectral Index  \\
 $T-T_{trig}$ (s)     &            &              &		   &  	          \\
\hline
 28-50           &  BAT & 15-150 keV       & 	-     &  -   \\
 50-60           &  BAT & 15-150 keV     & 	-     & $0.3\pm0.2$ \\
\hline
28-75		 & TAROT  &  $R$ -band    & $2.4\pm0.2$    &     -               \\
75-26000	 & TAROT+MITSUME+RTT  &  $R$ -band      & $1.18\pm0.05$  &     -                \\

\hline
85-200		 & UVOT  &   $V$  -band    & $1.2\pm0.4$    &     -               \\
\hline
A(87-150 s)       & XRT/WT  & 0.3-10 keV  &  $3.8\pm0.5$   & $1.0\pm0.2$ \\
B(150-1000s)      & XRT/WT+PC& 0.3-10 keV   &  $1.25\pm0.15$ & $1.2\pm0.2$ \\
C(4350-6800s)     & XRT/PC  & 0.3-10 keV    &  $0.6\pm0.2$   & $1.3\pm0.2$   \\
D(9930- 185650s)  & XRT/PC  & 0.3-10 keV    &  $1.4\pm0.1$   & $1.1\pm0.1$   \\
\hline
\end{tabular}
\label{xaft}
\end{table*}

\section{The host galaxy of GRB 060111B}
\label{res2}

\subsection{Imaging}
\label{sub-imaging}

Two deep exposures have been taken with the Canada-France-Hawaii Telescope (CFHT\footnote{Based on observations from director discretionary time obtained at the Canada-France-Hawaii Telescope (CFHT) which is operated by the National Research Council of Canada, the Institut National des Sciences de l'Univers of the Centre National de la Recherche Scientifique of France, and the University of Hawaii}) 5.3 months and 2.3 years after the burst with the $R$ and $H$ band filters, respectively. We detect a persistent source in the $R$-band, consistent with the afterglow position of \grb11, with magnitude $R=24.8\pm0.4$. The host is $6.3''$ off-axis 
from a nearby, bright spiral galaxy. From the positions derived from the relative astrometry performed on both the UVOT and CFHT images, we found that the afterglow and the host galaxy have 
an offset of less than 0.2 arcseconds (Fig. \ref{fig:f9}). 
The \grb11 field was also observed in the $H-$band for an exposure time of 30 minutes with the WIRCAM wide-field NIR camera at the CFHT on May 18, 2008  (2.3 years after the burst). Data were preprocessed and calibrated by ``I'iwi'' pipeline at CFHT, and  stacked with the SWARP software. We could not detect any source at the location of the burst down to the limiting magnitude of 23.0 (22.0) for a signal to noise ratio of 1.5 (3.0) (AB magnitude system). 

From the CFHT image and considering that at the epoch of its 
exposure the afterglow has definitively faded away, it is clear that \grb11 
occurred within a faint galaxy.

\subsection{Spectroscopy}

\begin{figure}
   \centering
   \includegraphics[width=8cm,angle=0]{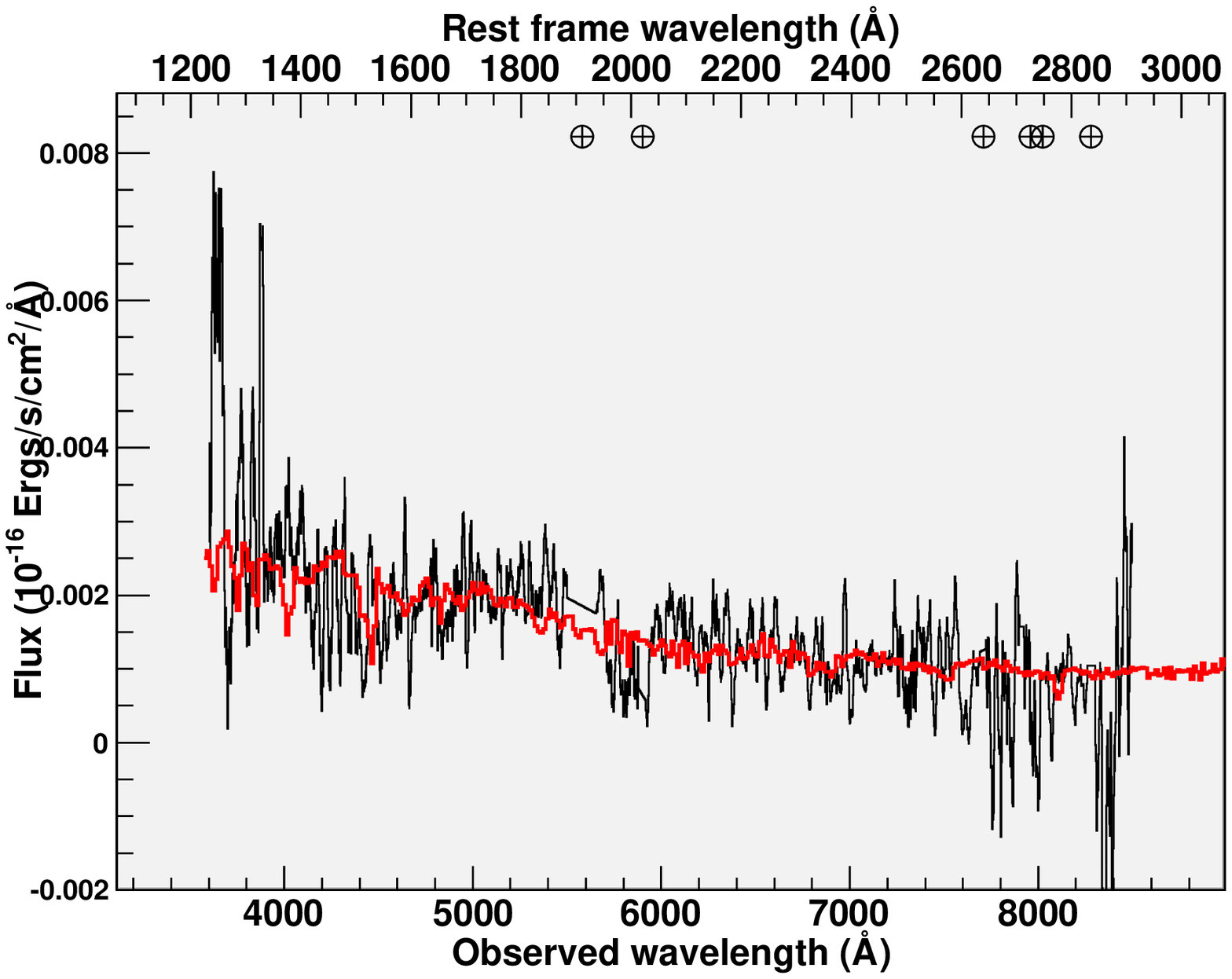}
   \includegraphics[width=8cm,angle=0]{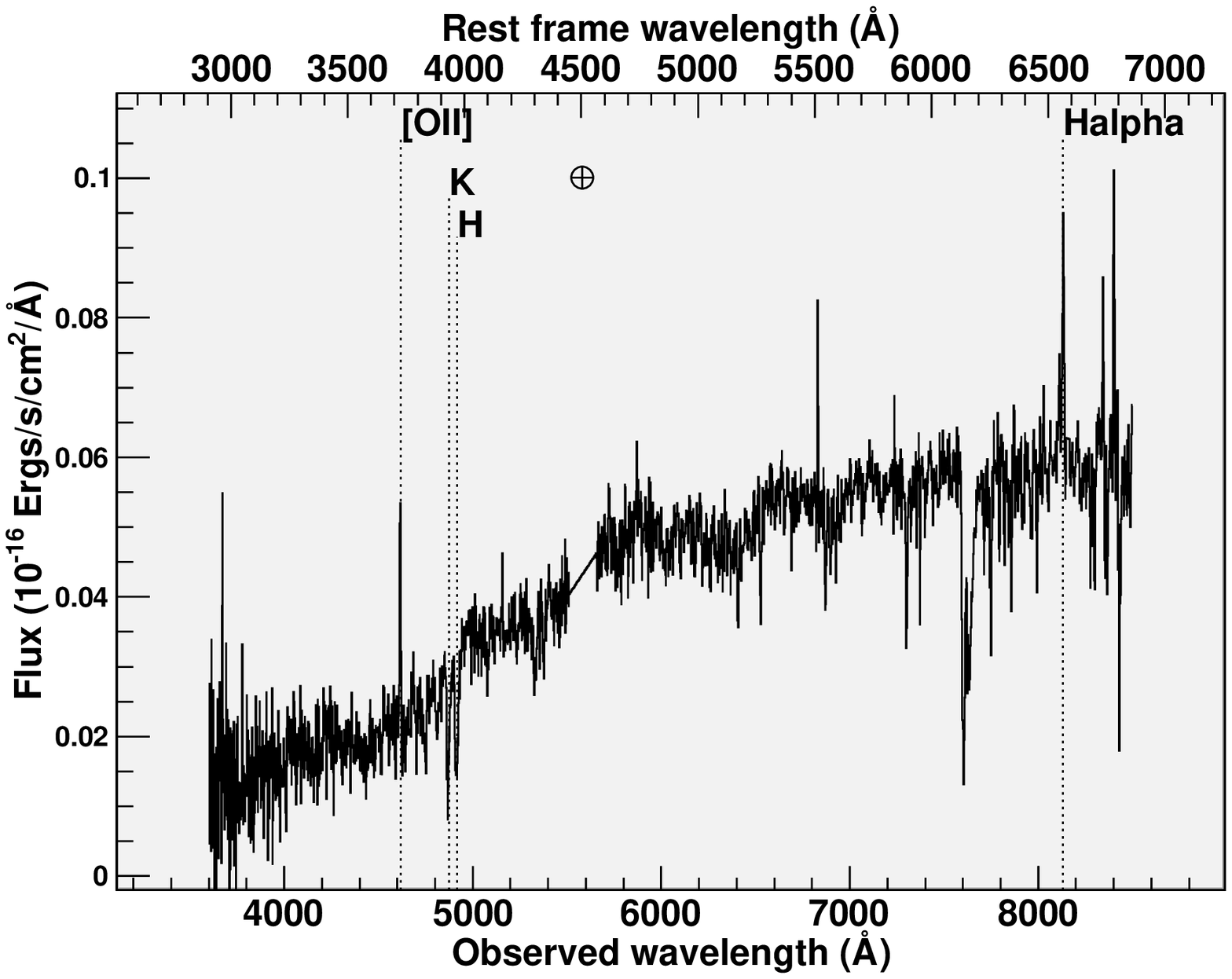}
      \caption{{\it Upper panel}: optical spectrum of the host. The red curve is the template model.  
{\it Bottom panel}: optical spectrum of the nearby galaxy taken with the Keck Telescope.
    }
         \label{fig:spectrum-host}
   \end{figure}

A spectroscopic observation of the host galaxy has been performed the by LRIS instrument on the Keck I telescope on 2006 July 26 (UT). The 600/8500 grism and 600/4000 grating with the 560 dichroic were used in order to allow the covering of a wide wavelength domain (between 3800 and 8500 \AA). The slit width was fixed to 1.5 arcsec and it was oriented in order to cover both the GRB host galaxy and the bright nearby galaxy (see \S 4.3). The total integration time was $2\times2400$ s. Due to the position of the object in the sky, the airmass was relatively large during the observation, about 1.6. Data were reduced using the LRIS rutines within the IRAF package. The two arms, Blue and Red, were reduced separately. In particular, the over-scan region was removed and the bias was subtracted by using the {\it lrisbias} task, which manages the dual amplifier (two-amp) readout mode.  Since the two LRIS arms do not have exactly the same spectral resolution, the two spectra were therefore rebinned to the same wavelength resolution of 1 \AA~  and finally combined together to produce the final spectrum. The junction region between the two arms, around 5800 \AA, presents anomalous features that were cleaned `manually' (Fig. \ref{fig:spectrum-host}). 

No clear emission or absorption lines are visible in the spectrum. No feature is compatible with a galaxy located close to, or at the same redshift than, the nearby large galaxy (see below). By applying a simple boxcar filter over 15 pixels, a weak continuum is observed, which is compatible with the one from a starburst spiral with $0.39<E(B-V)<0.50$ \citep{Kinney1996} at a redshift $z\sim 1.9$.

\subsection{The bright nearby galaxy}
\label{obs-host-bgal}

A much brighter galaxy is present in our imaging as well, centered at an angular offset of only 4.5 arcseconds from the GRB host galaxy. 
Astrometry based on CrAO $R-$band observations taken 16 days after burst and USNO-A2.0 reduction located the bright galaxy at coordinates (J2000) R.A.=19$^h$ 05$^m$ 43.06$^s$, Dec.=+70d 22$'$ 29.7$''$  with uncertainty of $0.2''$ (Fig. \ref{fig:f9}). 
We performed first photometric and then spectroscopic observations of this galaxy in order to check whether any connection with the host (e.g. a satellite galaxy) and/or the burst itself (e.g. a halo burst as if it originated from an old binary system) could exist. 
Photometric observations have been performed at the NOT telescope about four and 
eight months after the burst. $U$, $B$, $V$, $R$, $I$ and $J$ magnitude are quoted in Table \ref{photom-gal}. The best  photometric redshift obtained with HyperZ code \citep{Bolzonella2000} is $z=0.22^{+0.06}_{-0.07}$.
Its type is compatible with an early spiral galaxy, Sb. From $J$ band images taken at NOT with WFcam, we find marginal 
evidence of two spiral arms. The angular scale at the measured redshift is 3.76 kpc/arcsec, making the size of this galaxy close to $5.5''\times3.76$=20.7 kpc. 
Hereafter, we assumed a flat cosmology with $H_0=71$ km s$^{-1}$ Mpc, $\Omega_{\Lambda}=0.73$, $\Omega_m=0.27$ (Wright 2006).  
From spectroscopic analysis performed of the Keck spectrum (see previous section), 
the galaxy displays several emission and absorption lines (Fig. \ref{fig:spectrum-host}), allowing redshift measurement of $z=0.239$ (1 $\%$ accuracy).

   \begin{table}
      \caption[]{Photometry of the nearby galaxy (not corrected for Galactic extinction).}
         \label{photom-gal}
     $$ 
         \begin{tabular}{rrrr}
            \hline
            \noalign{\smallskip}
Filter & magnitude & $\Delta$mag & Exposure  \cr
       &           &             &   (s)       \cr
            \noalign{\smallskip}
            \hline
            \noalign{\smallskip}
U  & 23.34 & 0.3 & 1500 \cr
B  & 22.54 & 0.06 & 810 \cr
V  & 21.02 & 0.03 & 360 \cr
R & 20.24 & 0.02 & 360 \cr
I$_{\mathrm int}^{\mathrm{a}}$  & 19.70 & 0.03 & 180 \cr
J                 & 17.63 & 0.05 & 2880 \cr
            \noalign{\smallskip}
            \hline
         \end{tabular}
     $$ 
\begin{list}{}{}
\item[$^{\mathrm{a}}$] Filter I$_{int}$ is centered at $\lambda$=7970\AA (FWHM=157 \AA) \#12 interference I filter
more details in http://www.not.iac.es/instruments/filters/filterlist.html
\end{list}
   \end{table}

\subsection{Limits on the SN component}

No detection of any associated supernova (SN) down to $R=25.1$ was observed in the ZTSh 
telescope observations taken 16 days after the GRB (Tab. \ref{tab-opt}). 
Assuming as GRB-SNe templates the cases of SN1994I (GRB 021211, \citeauthor{Dellavalle2004}  2004) 
and SN1998bw (GRB 980425, \citeauthor{Galama1998} 1998) and considering 
a host galaxy dust extinction in the  $V$  band of about $\sim1$ the lack of any SN detection in our images implies a redshift lower limit of about 0.4 for a SN1994I-like event and of 
about 0.7 for a SN1998bw-like one.

\subsection{Redshift estimate}

In this section we attempt to estimate the distance of \grb11 
following some considerations based on our multi-band observations of both the host and the afterglow.

The host galaxy spectrum shows a weak continuum that is compatible with a starburst spiral with $0.39<E(B-V)<0.50$ \citep{Kinney1996} at a redshift $\sim 1.9$. Unfortunately, no emission or absorption lines are present (Fig. \ref{fig:spectrum-host}).  The two most common lines observed in long GRB host galaxies are the [OII] line and the H$\alpha$. Although the lack of these lines cannot give an absolute restriction to the redshift range, the lack of [OII]($\lambda$ 3727 \AA) emission line may imply ${\mathrm z} > 1.5$ while the lack of $H\alpha$($\lambda$ 6563 \AA) emission line may imply ${\mathrm z} > 0.4$. 

The best-fit redshift from the afterglow SED was $z=2$, consistent with the result obtained from fitting the host spectral continuum. We also used the X-ray light curve to estimate the distance to this burst:  following \cite{Gendre2008c} we estimate $z=2.2\pm1.0$, consistent with the above results. 
Finally, comparing the host galaxy apparent magnitude of $R\sim25$ with the typical magnitudes observed for a sample of long GRBs hosts, we find that it is likely at $z\sim1-2$ (cf. for instance Figure 9 of Berger at al. 2007), again consistent with previous findings. 

We further investigate on the distance scale of this burst by comparing the energetics and the host galaxy properties obtained assuming different redshifts, with the typical values observed for long GRBs. In Table \ref{radius} we have computed the size of the host (assuming an angular size of $1''$, corresponding to the seeing of CFHT observation) and its absolute magnitude and the burst energetics at several different  redshifts. The burst energetics was estimated with the isotropic equivalent  energy $E_{iso}$ that we have evaluated through the observed 100-1000 keV fluence $S=(5.6\pm0.8)10^{-6}$ erg cm$^{-2}$ (see \S 2.1) as  $S\times 4\pi D_L^2/(1+z)$ where $D_L$ is the luminosity distance.  
The absolute magnitudes have been computed from the 
observed $R$ magnitude, considering that for $z\ge0.5$ it samples the rest 
frame $B$ filter magnitude while for $z\ge1$ we assumed $U-B\sim0.75$ mag 
as typical $U-B$ color of starburst galaxies \citep[e.g.][]{Coil2007,Berger2009}.

\begin{table}[ht!]
\centering
\caption{Host galaxy and burst properties at different redshifts.}
\begin{tabular}{ccccccc}
\hline
$z$    &  d$^{\mathrm{a}}$ &   $M_B^{\mathrm{b}}$  &    $E_{iso}$   &  $E_{peak,i}$ & $\theta_j^{\mathrm{c}}$ \\
     &  kpc&  mag &   $10^{52}$erg &    MeV      & deg   \\
\hline
0.24 & 3.76  &   -15.4 & $0.07\pm0.01$ &  $0.6\pm0.3$ & 4.5\\ 
1    & 8.0   &   -18.4 & $1.5\pm0.2$   &  $0.9\pm0.5$ & 2.6\\ 
2    & 8.5   &   -20.3 & $5.8\pm0.8$   &  $1.4\pm0.8$ & 1.9\\ 
3    & 7.8   &   -21.5 & $12\pm2$     &  $1.8\pm1.0$  & 1.5\\
\hline
\end{tabular}
\begin{list}{}{}
\item[$^{\mathrm{a}}$] host size
\item[$^{\mathrm{b}}$] absolute magnitude of the host
\item[$^{\mathrm{c}}$] jet opening angle assuming that the temporal break observed in the X-ray light curve at a$\sim6$ ks has a jet origin 
\end{list}
\label{radius}
\end{table}

\begin{figure}
   \centering
	\includegraphics[width=8cm]{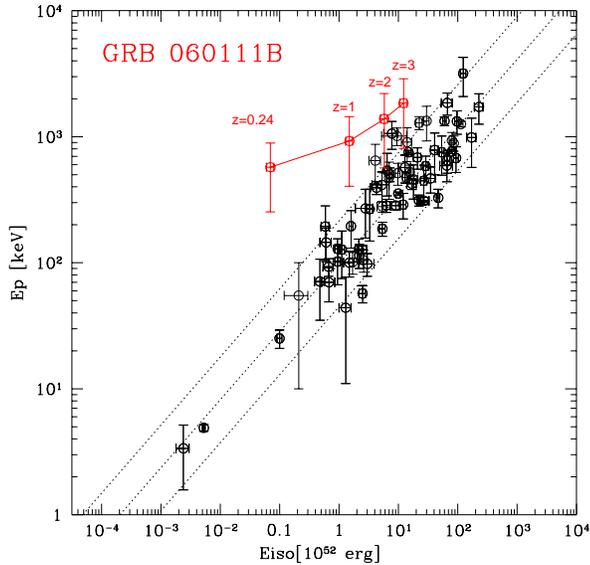}
      \caption{GRB 060111B is plotted (in red) in the $E_{peak}$vs$E_{iso}$ plot at four different redshifts: $z=0.24$,1,2 and 3, the latter being the maximum redshift compatible with our observations (see \S 5.2). Other GRBs data and the best fit correlation whithin 2 $\sigma$ are taken are from \cite{Amati2008}. }
         \label{amati}
   \end{figure}

Long GRB host luminosities cover a wide range of values, most of them  
between $M_B=-17$ and $M_B=-22$ mag, with a peak at about $M_B\sim-18$ mag 
\citep{Berger2007} while the typical host sizes is 
$\sim1.7$ kpc \citep{Wainwright2007}. 
Considering that the host angular sizes quoted in Table \ref{radius} should be considered as 
an upper limit due to the seeing at CFHT at the epoch of the observations, 
the host size of \grb11 is consistent with typical long GRBs host sizes at all the tested  redshifts while its absolute magnitudes would be too faint at the redshift of the nearby galaxy.  
From burst energetics considerations, Table \ref{radius} shows that at $z=0.24$, \grb11 would be an outlier in the Amati $E_{peak,i}$ versus $E_{iso}$ correlation (Fig. \ref{amati}). We note however that even at higher redshift \grb11 seems to deviate from the correlation\footnote{This may be the reason why the pseudo-redshift (Atteia 2004) computations provides the uncomfortably high value of $pz\sim8.4$.}.
Other long GRBs detected by BATSE are inconsistent with the Amati correlation \citep{Nakar2005,Band2005} and \grb11 may fall into this category (Fig. \ref{amati}). However, the large uncertainties affecting the peak energy measure  prevented us to reach any firm conclusion.

All of these considerations argue against any relation with the nearby, bright galaxy. Consistent with this conclusion is the lack of any supernova detection that imposes a redshift larger than $z\sim0.4-0.7$. For the latter, we assumed an observed magnitude of R$\ge$25.1 mag for the SN; that is, the same brightness as the source detected in the ZTSh telescope images taken 16 days after the burst (see \S 4.4). 

An upper limit on the redshift can be obtained from the lack of evidence of Lyman alpha cut-off due to the absorption from neutral hydrogen in the intergalactic medium in the spectral continuum of the host galaxy and by the $B-$band detection, providing ${\mathrm z} < 3$.  
 
Summarizing the above considerations, there is evidence against a possible relation of \grb11 with the nearby, bright galaxy and independent estimates converge to $z\sim1-2$. Having GRB 060111B at this redshift would result in having a standard long GRB host galaxy.

   \begin{figure}
   \centering
   \includegraphics[width=8cm]{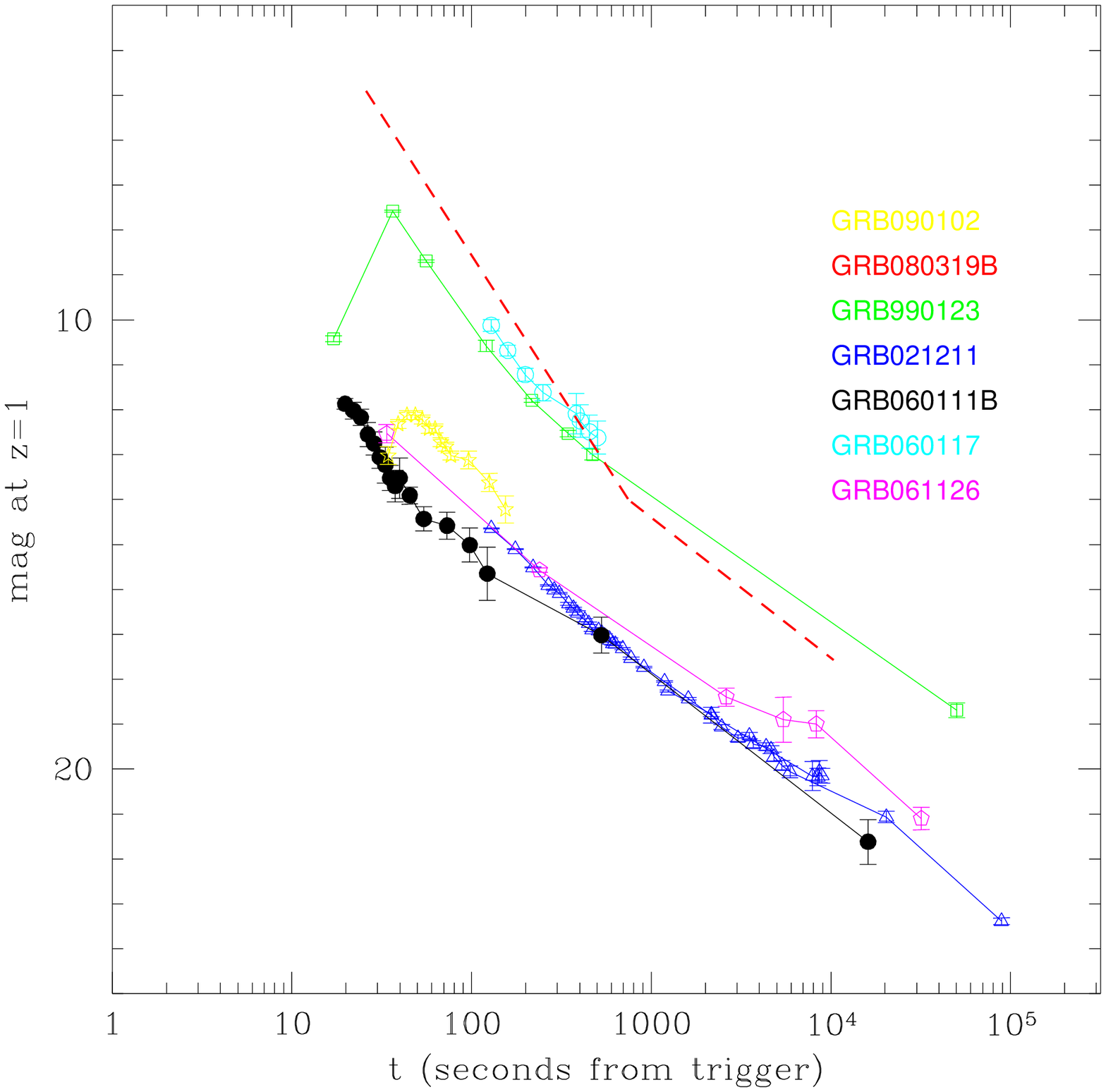}
      \caption{Early optical emission light curves as if all bursts were at $z=1$ of GRB 060111B (filled circles), GRB 990123 (open squares, \citeauthor{Akerlof1999} 1999), GRB 021211 (open triangles, \citeauthor{Fox2003} 2003, \citeauthor{Li2003} 2003), GRB 060117 (open circles, \citeauthor{Jelinek2006} 2006), GRB 061126 (open pentagons, \citeauthor{Perley2008} 2008), GRB 080319B (dashed line, best fit from \citeauthor{Racusin2008b} 2008b) and GRB 090102 (cyan pentagons, \citeauthor{Klotz2009} 2009). For these GRBs the Reverse Shock model has been claimed to explain the prompt optical afterglow (see Table \ref{lc_opt}).}
         \label{test}
   \end{figure}

\subsection{Outflow collimation}
\label{discuss-prompt}

Using our redshift estimates above, we can compute the possible collimation of the outflow $\theta_j$  
assuming a jet origin of the temporal break $t_b$ observed in the XRT light curve at about 6 ks.  The angle estimations computed assuming a constant ISM density of $n\sim5$ cm$^{-3}$ and radiative efficiency $\sim0.2$ are presented in Table 5. It occurs that the jet opening angle is within the distribution of angles of sample GRBs with well defined jet break in X-ray light curve \citep{Racusin2008b}. Alternatively, if the jet break epoch is beyond the end of XRT observations, that is $t_b>200$ ks, then $\theta_j > 6$ deg ($z\le2$)  which is marginally consistent with the distribution. Thus we can support the break at $t_b\sim6$ ks as a real jet break. This result is in line with the anticorrelation found between the X-ray afterglow shallow phase durations and the redshift \citep{Stratta2009}, that, if interpreted as a selection effect, may reflects an anticorrelation of the X-ray shallow phase duration with the burst energy. Indeed, the latter is expected if the end of the shallow phase originates from a collimated outflow. The burst energy corrected for collimation at $z=1-2$ would be $E_{\gamma}\sim1-2\times10^{49}$ erg.

\section{The nature of the prompt optical emission}
\label{discuss-prompt}

The nature of the prompt optical emission of \grb11 has been already investigated in past works \citep[e.g.][] {Klotz2006,Wei2007}.  A reverse shock (RS, e.g. Meszaros \& Rees 1994,1997) emission has been invoked for this burst by \cite{Klotz2006}, primarily based on the TAROT observations of very steep early-time optical decay ($\alpha=-(2.4\pm0.1)$ in $R-$band, see \S 3.1). Further evidence includes the observed lack of temporal  correlation between the simultaneous BAT and TAROT emissions and the similarities of its optical light curve with GRB 990123 and GRB 021211 (for which a reverse shock origin has been proposed to explain their optical light curves). 
A decay index of $\sim-2$ has generally been considered as the evidence of a RS origin. Some concerns however have been addressed to the unknown RS zero-time that, if different than the trigger time, would make the real decay rate of the emission different from those measured, leading to different conclusions \citep{Panaitescu2009}. A simultaneous spectral and temporal analysis would be the best way to investigate the nature of the observed optical radiation. 
Unfortunately, for \grb11 we do not have any spectral information in the optical regime during the steep decay. 
In this work we further investigate on the nature of the prompt optical emission by exploring its optical-to-gamma ray spectral energy distribution. We also compare \grb11 with other GRBs showing early optical emission with similar temporal properties. We then attempt to constrain the physical parameters of this burst using our combined early-through-late-time observations, including our constraints on the redshift and on the host-galaxy dust extinction. 

Thanks to our multi-band analysis we were able to confirm the presence of non-negligible extinction from dust in the burst environment. By correcting the TAROT data for the estimated amount of host galaxy dust extinction, we could evaluate the intrinsic optical-to-gamma-ray flux density ratio during the prompt emission. The obtained values are orders of magnitude above unity, quite unlike most GRBs with simultaneous optical counterpart which have been reported to date (see \citeauthor{Yost2007} 2007). Most importantly, the optical flux density is well above the extrapolation obtained from the best fit spectral model in the gamma-ray regime. This result, as well as the smooth, fast decay of the optical light curve simultaneous to the strong gamma-ray flux variability, enabled us to firmly exclude a common emission mechanism at high and low energies.  
To compare \grb11 with other GRBs with "steep-to-shallow" decay observed during early optical emission, we plotted   the optical magnitudes as a function of time as if all bursts were at $z=1$ (Figure \ref{test}). To compute the expected magnitudes at $z=1$ we simply apply the distance modulus definition without taking into account any k-correction and the host galaxy extinction correction since for almost all the plotted bursts we have no color information during the early optical emission and very uncertain or no host galaxy extinction measurements. For GRB 080319B,  given the huge amount of optical data available, we decided to plot only the well constrained best fit light curve model \citep{Racusin2008b} in the period of interest for our study (that is, when the flux features the steep-to-shallow decay). Inspection of the figure demonstrates the strong similarities in the initial steep decay indexes (all within $\sim2$ and $\sim2.5$, see Tab. \ref{lc_opt}).
This evidence possibly indicates the presence of a common starting time of the
observed prompt optical emissions that, if coincident with the trigger time,
provides good consistency with the RS predicted decay slopes.

In the context of the internal-external shock model \citep[e.g.][]{Sari1997a}, the RS emission mechanism is a consequence of the external shock that the outflow produces with the surrounding interstellar matter, along with a forward shock (FS) that generates the afterglow emission. As for the FS, the RS accelerates electrons in the outflow into a power law energy distribution with index $p$ \citep[e.g.][]{Sari1998,Sari1999}. The light curve at a given frequency depends on the temporal evolution of the injection frequency $\nu_m$, the cooling frequency $\nu_c$ and the synchrotron peak flux $F_{\nu,max}$ \citep[e.g.][]{Sari1998}.  Kobayashi (2000) derived a simple analytic solution of the RS light curve in two limits: the thick shell limit, when the RS becomes relativistic crossing the outflow and the thin shell limit, when the RS remains Newtonian. From an observational point of view the thick shell limit predicts that the RS peaks during the high-energy burst while a RS peaking after the end of the burst is predicted from the thin shell limit case. The combinations of the RS and FS light curves have been classified by \cite{Zhang2003} as of type I,II and III depending on RS and FS relative peak times and peak flux intensities.  \cite{Klotz2006} pointed out that the optical prompt emission of \grb11 can be classified as type II \citep{Zhang2003} where the RS emission is dominating over the peak of the FS component after the FS peak time.


In order to compare observations with theoretical predictions, we first attempt to constrain the fireball initial Lorentz factor $\Gamma$ from the time at which we observe the prompt optical decay, \citep{Sari1997b,Zhang2003}. Since the optical light curve is already fading at $t=28$ s after the trigger, the rest frame deceleration time is at $t_d \le t/(1+z)$; that is, the fireball has given the ISM an amount of energy comparable to its initial energy at  $t_d=[3E/32\pi n m_p c^5 \Gamma^8]^{1/3}\le t/(1+z)$. Assuming an interstellar matter density of $n\sim 5$cm$^{-3}$ and a redshift of $z=1-2$ and estimating the energy of the shell $E$ with $E_{iso}$, we find $\Gamma\ge$260-360.  This result is consistent with past estimates of the fireball Lorentz factor of about a few hundreds \citep[e.g.][] {Molinari2007,Gomboc2008,Ferrero2008}. 


As already mentioned by \cite{Klotz2006} \grb11 falls in the thick shell limit since the optical prompt emission is already fading before the end of the prompt, high-energy burst. In this limit, the deceleration time should coincide with the time at which the RS crosses the shell \citep[e.g.][]{Kobayashi2000,Zhang2003} that is estimated with the high-energy burst duration. 
We use $T_{90}\sim25$ s observed in the Suzaku/WAM energy range ($100-1000$ keV) to estimate the burst duration. The injection frequency, the cooling frequency and the peak power at the shock crossing time for the thick shell limit are \citep{Sari1999,Kobayashi2000}: 

\begin{equation}
\nu_m \sim 7.3 \times 10^{14}\epsilon_{e,0.6}^2 \epsilon_{B,0.01}^{1/2} n_5^{1/2}\Gamma_{300}^{2} Hz
\end{equation}

\begin{equation}
\nu_c \sim 4.5 \times 10^{15} \epsilon_{B,0.01}^{-3/2} E_{52}^{-1/2} n_5^{-1} (T_{100}/(1+z))^{-1/2}  Hz
\end{equation}

\begin{equation}
F_{\nu,max} \sim 130 D_{L,28}^{-2} \epsilon_{B,0.01}^{1/2} E_{52}^{5/4} n_5^{1/4} \Gamma_{300}^{-1} (T_{100}/(1+z))^{-3/4} mJy 
\end{equation}

where $D_{L,28}=D_L/10^{28}$ cm and $E_{52}=E_{iso}/10^{52}$ erg are the luminosity distance and the outflow energy computed at $z=2$ (Table \ref{radius}),  $\epsilon_{B,0.01}=\epsilon_B/0.01$ and $\epsilon_{e,0.6}=\epsilon_e/0.6$ are the fractions of the burst total energy that goes into the electrons and magnetic field respectively, $n_5=n/5$  cm$^{-3}$ is the interstellar matter density, $T_{100}=T_{90}/100$ s, and $\Gamma_{300}=\Gamma/300$ is the outflow Lorentz factor.  Assuming the specified values, from equations (1) and (2) we find that at the RS crossing time the observed R-band frequency ($\nu\sim 1.2 \times 10^{15}$ Hz in the rest frame), is between $\nu_m$ ($7.3 \times 10^{14}$ Hz) and $\nu_c$ ($4.5 \times 10^{15}$ Hz). 
Equation (3) shows that the RS peak flux density is expected at $\sim$130 mJy for $z=2$. 
From Figure 5 we see that at the beginning of the prompt optical decay, the flux density is $\sim10$ mJy. This value however is not corrected for the dust extinction in the host. The latter is highly uncertain but we find evidence that is not negligible (see \S 3.2). From our optical-to-X-ray best-fit obtained assuming a starburst extinction curve, we measure a rest frame UV extinction of about 2.5 mag that makes the corrected measured flux density consistent with the expected peak flux. This result indicates that the RS peak time should not be too far from the TAROT start observing time and that the outflow Lorentz factor is close to the estimated lower limit value. 


From (1) and (2) we obtain that $\nu_m<\nu_c$ at the RS crossing  time. Therefore, for the TAROT observations we can assume a slow cooling regime. For $\nu_m<\nu<\nu_c$ as observed, $F_{\nu}=F_{\nu_m}(\nu/\nu_m)^{-(p-1)/2}\propto t^{-\alpha}$. Since $\nu_m \propto t^{-73/48} \sim t^{-3/2}$ and $F_{\nu_m} \propto t^{-47/48} \sim t^{-1}$ \citep{Kobayashi2000}, we expect $\alpha=(3p+1)/4$ that is consistent with our measured decay index if $p=2.9\pm0.3$. We note that this value is somewhat higher than what derived from the FS component (see \S3.2). Different values of $p$ for the FS and the RS have been  invoked as a possible scenario also for GRB 061126 by \cite{Perley2008}.

It has been argued that the RS emission mechanism cannot be the dominating mechanism to explain the bright optical early emission due to the lack of this component in many other GRBs despite the expected high flux \citep{Roming2005}. We note that the prompt optical emission in GRB 060111B ends very early compared to the other events (Fig. \ref{test}), opening the possibility that this component may remain undetected in a significant fraction of GRBs. In addition, the magnetization of GRBs outflows plays a crucial role in suppressing the RS optical emission \citep[e.g.][]{Fan2008}. 

In a scenario alternative to the RS, the prompt optical emission may be produced by internal shocks  different from those responsible for the gamma-ray emission, which can be formed by collisions of different shells: these shells may be easily disrupted by other shells, thus explaining why bright optical emission is not common in GRBs \citep{Wei2007}. For \grb11, \cite{Wei2007} computed that the rest frame cooling frequency should be at energies $\nu_c\sim0.3-0.7$ keV, that is between the X-ray and the optical energy domains, at $z=1-2$, 28 s after the trigger and assuming a Lorentz factor of the merged shell of 800. The $\nu_c$ expected value at the shell crossing time is consistent with our observations assuming that $\nu_c$ remains at energies higher than the optical up to $t_{trig}+200s$ after which the decay indexes in the two energy domains are consistent among themselves (see \S 3.1). However, the $\nu_c$ theoretical estimation depends on the highly uncertain fireball Lorentz factor to the power of eight.  If for example we assume $\Gamma=300$, the cooling frequency would had been well below the $R-$band at $t_{trig}+28s$. For this reason and for the numerous pieces of evidence supporting the RS emission exposed above, we do not favor the internal shock interpretation. 

Finally we mention that the RS emission mechanism has been proposed to solve the difficulties of the `standard' forward shock model to explain the canonical (steep-flat-normal) evolution of the X-ray emission \citep{Uhm2007,Genet2007b}. GRB 060111B represents one of the few cases where both an optical flash and a `canonical' X-ray light curve \citep{Nousek2006} are observed, enabling to test the RS in both the energy ranges.  
For example, to reproduce the observed X-ray light curves a long-lived RS emission is required, which is difficult to reconcile with bright optical early emission \citep{Genet2007b}.

\begin{table}[ht!]
\centering
\caption{Best-fit decay indexes of the early optical emission for all GRBs with well monitored optical counterpart that shows a steep-to-shallow behavior (Fig. \ref{test}).}
\begin{tabular}{ccccccc}
\hline
GRB    & Filter & $\alpha_1$	 &   $\alpha_2$  &  $t_{break}$  & $z$ \\
	&	&		&		& 	(s)	&	\\
\hline
990123$^{(1)}$ &  $V$  & $\sim2.5$  & $\sim1.5$ & 700  & 1.60	 \\
021211$^{(2)}$ &  $R$  &$1.97\pm0.05$  & $1.01\pm0.02$ & 300	 &1.0\\
060117$^{(3)}$ &  $R$  & $2.49\pm0.05$  & $1.47\pm0.03$ & 300	 &1$^{\mathrm{a}}$\\
061126$^{(4)}$ &  $R$  & $2.0\pm0.3$  & $0.76\pm0.06$ & 780 & 1.1588\\
060111B &  $R$  & $2.4 \pm 0.2$  & $1.18\pm0.05$ & 75 & 2\\
080319B$^{(5)}$&  $R$  & $2.49\pm0.09$  & $1.25\pm0.02$ & 900&0.937  \\
090102$^{(6)}$&  $R$  & $1.83\pm0.13$  & $1.1\pm0.3$ & 400 &1.547 \\
\hline
\end{tabular}
\label{lc_opt}
\begin{list}{}{}
\item[$^{\mathrm{a}}$] For GRB 060117 since the redshift is not known, we imposed z=1.
\end{list}
References. (1) \cite{Akerlof1999}; (2) \cite{Holland2004}; (3) \cite{Jelinek2006}; (4) \cite{Perley2008}; (5) \cite{Racusin2008b}; (6) \cite{Klotz2009}. 
\end{table}

\section{Summary and conclusions}
\label{conclusion}

The presence of a bright optical emission during the prompt phase
makes \grb11 an extremely interesting burst and for this reason 
an extended multi-wavelength campaign has been performed in order 
to provide further constraints to the physical origin of the observed 
optical emission. Hereafter we summarize our main results and conclusions. 

Our late optical observations enabled us to identify the host galaxy of \grb11. The featureless host spectrum, the non detection of any associated supernova 16 days after the burst as well as the inferred host luminosities at different redshift, allowed us to put some constraints on the distance scale of \grb11. Independent estimates converge to $z\sim1-2$. 


From the afterglow SED we find that a non-negligible host galaxy dust extinction affects the observed flux in the optical regime. After correcting for this effect, the 
optical-to-gamma-ray flux ratio during the prompt strongly suggests a separate origin 
of the optical component from that one at high energies. This was already 
pointed out by \citep{Klotz2006} from the observed high energy prompt 
light curve not tracked by the simultaneous optical one. The similarities 
of the measured optical decay index of \grb11 with other observed prompt 
optical emissions suggest a common zero-time origin of the emission mechanism that, if coincident with the burst trigger time, makes the measured decay indexes consistent with reverse shock scenario. 

From the time at which we observe the TAROT light curve starting to decay, we estimated the fireball initial Lorentz factor \citep{Sari1997b} as  $\Gamma\ge$260-360 at redshift $z\sim1-2$, in agreement with past results obtained for other bursts. Assuming $\Gamma\sim300$, the observations are in good agreement with theoretical predictions from the reverse shock emission in the thick shell limit. 
In particular, we find evidence that the prompt optical emission is in the slow cooling regime with $\nu_m<\nu<\nu_c$ and that the peak flux is very near or coincident with the observed one corrected for host galaxy extinction. 

The very short duration of optical RS emission in GRB 060111B (about 30 sec in the restframe) raises the question of the true fraction of GRBs displaying this type of emission.  
\grb11 is among the few cases of a burst with bright early optical 
counterpart observed with good temporal resolution as well as a `canonical' 
Swift/XRT light curve in X-rays and provides an excellent test case for  
the reverse-shock emission mechanism in both the energy ranges.


\begin{acknowledgements}
The authors thank the anonymous referee for his/her useful comments and acknowledge Taka Sakamoto and Scott D. Barthelmy for 
making BAT data available to the community, and J.S. Bloom for obtaining the Keck data. 
BG is funded by a french CNES post-doctoral grant.
\end{acknowledgements}

\bibliographystyle{aa} 
\bibliography{grb5bi} 


\end{document}